\documentstyle[preprint,aps,amssymb]{revtex}
  \tightenlines
\begin{document}
\draft
\widetext

\title{\bf    Magnetic     pair-breaking    in    superconducting
  {Ba$_{1-x}$K$_x$BiO$_3$} investigated by magnetotunneling}
  \author{P.     Szab\'o,$^{1,2}$      P.     Samuely,$^{1}$
A.G.M.Jansen,$^{2}$  J. Marcus$^{3}$  and P. Wyder$^{2}$}
\address{$^1$Institute of Experimental Physics, Slovak Academy of
Sciences,         SK-04353         Ko\v{s}ice,         Slovakia.\\
$^2$Grenoble       High       Magnetic      Field      Laboratory,
Max-Planck-Institut     f\"{u}r     Festk\"{o}rperforschung    and
Centre National  de la Recherche Scientifique,  B.P. 166, F-38042
  Grenoble Cedex 9, France.
$^3$Laboratoire d'Etudes des Propri\'et\'es Electroniques
des Solides, Centre National de la Recherche Scientifique, B.P. 166,
F-38042 Grenoble Cedex 9, France. }
\date{\today}

\maketitle
  \widetext

\begin{abstract}

  The de Gennes and Maki  theory of gapless superconductivity for
  dirty  superconductors  is  used  to  interpret  the  tunneling
  measurements    on    the     strongly    type-II    high-$T_c$
  oxide-superconductor  Ba$_{1-x}$K$_x$BiO$_3$  in  high magnetic
  fields up to  30 Tesla. We show that  this theory is applicable
  at  all temperatures  and in  a wide  range of  magnetic fields
  starting from  50 percent of  the upper critical  field $B_{c2}$.
In this magnetic field range
the  measured  superconducting  density  of  states  (DOS)  has
the simple energy dependence as predicted by de  Gennes
from which
the temperature dependence of the pair-breaking parameter
  $\alpha(T)$,   or $B_{c2}(T)$,    has   been  obtained.  The
deduced temperature dependence  of $B_{c2}(T)$ follows the
  Werthamer-Helfand-Hohenberg  prediction  for  classical type-II
  superconductors   in   agreement   with   our  previous  direct
  determination.
  The amplitudes of the  deviations in the DOS depend on the magnetic field
  via the spatially averaged superconducting order  parameter
  which has a square-root dependence on the  magnetic field.  Finally, the
  second   Ginzburg-Landau  parameter   $\kappa_2(T)$  has   been
  determined from the experimental data.

\end{abstract}




\section{Introduction}

  The  temperature   dependence  of  the   upper  critical  field
  $B_{c2}$  in several  new classes  of superconductors  attracts
  much  attention as  it reveals  very unusual  behavior. For the
  conventional  type-II superconductors  $B_{c2}$ shows  a linear
  increase   below  $T_c$   and  a   saturation  at   the  lowest
  temperatures in  agreement with the  theoretical predictions of
  Maki   \cite{maki}   and   Werthamer,   Helfand  and  Hohenberg
  \cite{whh}.  In   the  case  of   certain  new  superconductors
  $B_{c2}(T)$ has  a positive curvature practically  in the whole
  temperature range.  As examples of this  anomalous behavior we
  mention   the   borocarbides   \cite{borocarb},   the   organic
  superconductors  \cite{organics},  the  high-$T_c$  bismuthates
  {Ba$_{1-x}$K$_x$BiO$_3$}   \cite{klein1},  and   in  the   most
  pronounced way the cuprates \cite{cuprates} where in some cases $B_{c2}(T)$
  even  diverges   at  very  low   temperatures.
  Several  scenarios  have  been  proposed  to
  explain this  anomalous behavior. Among others,  one can find
  a bipolaron model \cite{alexandrov93}, an unconventional normal
  state   \cite{dias94},   a   strong   electron-phonon  coupling
  \cite{marsiglio87},   and the  presence   of  inhomogeneities   and
  magnetic impurities \cite{ovchinnikov96}.

  In   the  above   mentioned  cases,  magnetotransport   or
  magnetization measurements were  employed for the determination
  of $B_{c2}(T)$. In strongly type-II superconductors a magnetization measurement
near $B_{c2}$ can  be very difficult  because of its
  extremely  small value.  Moreover, because depinned  vortices in
  the  liquid   or  solid  state  cause   a  finite  dissipative
  resistance before  reaching the full  transition to the  normal
  state,  the  complexity  of  the  $B  -  T$  phase  diagram  in
  high-$T_c$'s    \cite{blatter94}    undermines    any    direct
  determination of the upper critical field from magnetotransport
  data.  Recently, the $B_{c2}(T)$ dependencies
  have been determined  for high-T$_c$ superconductors
  by  non-dissipative  experimental  methods.  Carrington  et al.
  \cite{carrington}   have   shown,   that  magneto-specific-heat
  measurements    in the
  overdoped  Tl-2201 cuprates    yield in the high  temperature region
  a different curvature  of $B_{c2}(T)$ as compared to  that determined
  from   ac-susceptibility  or  from  magnetotransport.  Blumberg et al.
  \cite{blumberg}  determined the  upper critical  field from the
  electronic  Raman  scattering  in  the  same  cuprates and also
  obtained a conventional temperature  dependence with a negative
  curvature   and   saturation    at   low   temperatures.
  In   our  previous
  experimental  work elastic tunneling
  was used  as a direct  tool to infer  the upper critical  field
  $B_{c2}(T)$  in  the  high-$T_c$  oxide Ba$_{1-x}$K$_x$BiO$_3$  \cite{samuely98}.
  Such a method  is based on a measurement  of the very fundamental
  superconducting density of states (DOS), where the superconducting
  part of the $B-T$ phase diagram is defined
  from the  non-zero value of  the superconducting order  parameter
  ($\Delta \ne 0$).

  In the present paper  the tunneling characteristics measured on
  Ba$_{1-x}$K$_x$BiO$_3$ in the mixed state at very high magnetic
  fields are  discussed in the  frame work of  the de Gennes  and
  Maki theory  of the gapless  superconductivity \cite{degennes1,maki2}.
  We show that the theory is applicable at all
  temperatures below $T_c$ and in a wide range of magnetic fields
  below  the  upper  critical  one. In  the  experimentally
  measured  tunneling  conductances  (proportional to the spatially  averaged
  superconducting DOS) at high fields only the  amplitude of the deviations
  from  the  normal-state  tunneling  conductance depends
  on the  applied  magnetic   field.
  The energy dependence of these deviations is completely controlled by  the
  temperature.  The  renormalized  tunneling-conductance  traces
  reveal a  simple scaling behavior. This  allows us to determine
  the  temperature  dependence  of  the  pair-breaking  parameter
  $\alpha  (T)$ and  the  upper  critical field  $B_{c2}(T)$. The
  fitted  $B_{c2}(T)$   follows  the  Werthamer-Helfand-Hohenberg
  prediction for classical type-II superconductors and is also in
  agreement    with    our    previous    direct    determination
  \cite{samuely98}.  The   analysis  of  the   amplitude  of  the
  tunneling  conductance at  different magnetic  fields enables us to
  determine
  the  magnetic  field  dependence  of  the  spatially
  averaged   value $\bar{\Delta}(B)$   of   the   superconducting  order  parameter
  at   different   temperatures and the
  temperature dependence of  the second Ginzburg-Landau parameter
  $\kappa_2(T)$.

\section{Theory}

  The  behavior of  type-II superconductors  in the  presence of
  external magnetic fields has been described on the basis of the
  phenomenological   Ginzburg-Landau   (GL)   theory   \cite{GL}.
  Abrikosov \cite{abrikosov} showed that a type-II superconductor
  exhibits  a mixed  (vortex) state,  in which  the magnetic flux
  penetrates  the sample  in the  form of  quantized flux  lines.
  These results  of the GL  theory have validity  in a restricted
  temperature  region  near  the  transition temperature $T_{c}$,
  where  the lower  and  upper  critical fields  ($B_{c1}(T)$ and
  $B_{c2}(T)$)   are  both   much  smaller   then  their   values
  $B_{c1}(0)$ and  $B_{c2}(0)$ at zero  temperature. An important
  generalization  of  the  GL  theory   has  been  made  by  Maki
  \cite{maki3} and de Gennes \cite{degennes} in the case of dirty
  superconductors (where  the mean free path  $l$ is much smaller
  then the BCS superconducting coherence length $\xi_0$) extending this
  theory  to arbitrary  temperatures when  the magnetic  field is
  close to $B_{c2}(T)$. Maki has shown, that Abrikosov's original theory
  is applicable at all temperatures in the dirty limit, if in the
  expressions  for  the   temperature  dependencies  of  critical
  magnetic field and magnetization  two temperature dependent GL
  parameters $ \kappa _i$ are introduced
\begin{eqnarray}
\kappa        _{1}        &=&B_{c2}(T)/\sqrt{2}B_{c}(T),        \\
-M   &=&\frac{B_{c2}   -  B}{\mu_0 \beta  (2\kappa  _{2}^{2}-1)},
\end{eqnarray}
  where  $M$ is  the magnetization  and $\beta  $  a geometric
  constant of the vortex lattice.  The first GL parameter $\kappa
  _{1}(T)$, the ratio of the upper
  critical  field $B_{c2}(T)$  and the  thermodynamic critical field $B_c(T)$,
  is a  slowly temperature dependent
  function with  a $20\%$ larger  value at $0$~K  than at $T_{c}$.
  The parameter $\kappa _{2}$ is  connected with the slope of the
  magnetization    curve   near    $B_{c2}$.   Caroli    et   al.
  \cite{caroli1} showed that $\kappa _{2}(T)  $ will be equal to
  $\kappa  _{1}(T)$  within  $2\%$   for  a  dirty  bulk  type-II
  superconductor.

  A magnetic field  applied to a dirty  superconductor breaks the
  (time-reversal)  symmetry  of the  Cooper  pairs  and  leads  to
  a finite lifetime $\tau _{k}$ of the condensed paired electrons
  with  the pair  breaking parameter  $\alpha _{k}  =\hbar /2\tau
  _{k} $.  This decay process  competes with a  natural growth of
  pairs associated with a  characteristic lifetime $\tau (T)$ and
  related parameter  $\alpha =\hbar /2\tau  $ which is  connected
  with   the  upper   critical  field   $B_{c2}$  according to
  \begin{equation}  B_{c2}=\frac{\alpha \Phi  _{0}}{\pi D\hbar },
  \end{equation} where  $\Phi _{0}$ is  the flux quantum  and $D$
  the  diffusion   constant.  Neglecting  the   effects  of  spin
  paramagnetism   and  spin-orbit   scattering  the   temperature
  dependence of  the pair-breaking parameter  $\alpha (T)$ is
  given by
  \begin{equation}
\ln(T_{c}/T)=\Psi \left(\frac{1}{2}+\frac{\alpha }{2\pi k_{B}T}\right)-\Psi \left(\frac{1}{2}\right),
\end{equation}
where $\Psi (z)$ is the digamma function.

  Magnetic   pair-breaking  in   superconductors  depresses   the
  superconducting order with the occurence of
  quasi-particle states inside the otherwise forbidden
  energy gap.  For sufficiently strong pair-breaking, near $B_{c2}$,
   a gapless superconducting state exists.
Superconductivity with  a finite value of
the superconducting order parameter (pair potential $\Delta$) exists
without a minimum excitation energy in the quasi-particle excitation
spectrum.
This gapless superconducting behavior can be clearly seen in the
calculations of Skalski et al. \cite{skalski} for
the density of states of the superconducting excitation spectrum
in the presence of pair breaking.
De Gennes  \cite{degennes} has derived a very  simple expression
 for the  density of  states $N(E,{\bf r})$ in dirty
  superconductors in  the gapless region for  small $\Delta $ in
  magnetic fields near $B_{c2}$, given by
\begin{equation}
N(E,{\bf r})=N_{N}(0)\left[1-\frac{\Delta ^{2}({\bf r}, H)}{2}\frac{\alpha ^{2}-E^{2}}{%
(E^{2}+\alpha                                         ^{2})^{2}}\right],
\end{equation}
  where  $N_{N}(0)$  is  the  DOS  at  the  Fermi  surface of the
  superconductor in the normal state. In the range of validity of
  Eq.~(5),   $\alpha  $ does not depend on magnetic field,
   but is a function   of    temperature   only.
   Via Eq.~(3),   the pair-beaking parameter $\alpha (T)$
   is related to $B_{c2}(T)$.
   The only  magnetic field  dependent parameter  in Eq.~(5)
   is $\Delta ({\bf  r},H)$. It  leads to  the important  conclusion that the
  energy dependence of $\ N(E, {\bf r})$ is fully
  controlled by temperature and not by magnetic field
  or $\Delta $. Via $\Delta $, the magnetic field  acts just as a scaling
  parameter for the amplitude of the deviations  of $N(E,{\bf r})$
  from $N_{N}(0)$.

  A direct     experimental    verification     of     gapless
  superconductivity in the presence of a pair-breaking perturbation
  can be  obtained by tunneling  spectroscopy \cite{degennes1,maki2}.
The differential tunneling   conductance  measured   on   a
  metal-insulator-superconductor  (N-I-S)  tunnel-junction is
  directly   proportional   to   the   spatially averaged   value   of  the
  superconducting DOS by
\begin{equation}
G(V)= \frac{(dI/dV)}{(dI/dV)_N}  = \int_{-\infty }^{\infty }\frac{N(E)}{N_{N}(0)}\left[\frac{%
\partial                f(E+eV)}{\partial                (eV)}\right]dE,
\end{equation}
  where $G(V)$ represents the tunneling conductance normalized to
  its  normal-state  value  and  the  brackets  in  the integral
  contain  the   bias-voltage   derivative   of   the   Fermi
  distribution function.  In  the case of a gapless
  DOS the normalized conductance  may be transformed into \cite{guyon}
\begin{equation}
G(V)=1+\frac{  \Delta  ^{2}}{8\pi  ^{2}k_{B}^{2}T^{2}}Re \left\{ \Psi  _{3}\left(%
\frac{1}{2}+a-ib\right)\right\},
\end{equation}
  where  $a=\alpha /2\pi  k_{B}T$, $b=eV/2\pi  k_{B}T$ and  $\Psi
  _{3}$  is the  second derivative  of the  digamma function.  In
  a geometry with the applied  magnetic field perpendicular to
  the planar surface of the junction barrier, the sample  is in the  Shubnikov mixed phase
  with the Abrikosov vortex structure. In a tunneling  experiment
  the spatial average  $\bar{N}(E)$ of the density of states over the sample surface
  is measured. Therefore, one has to consider in Eq.~(5) the spatially averaged
  order parameter $\bar{\Delta}$
  \cite{degennes1}.

  The magnetic  field dependence of  $\mid \bar{\Delta}\mid $  in
  the gapless region of  type-II superconductor near $B_{c2}$ has
  been calculated by Maki \cite{maki3} yielding
\begin{equation}
\bar{\mid \Delta \mid }^{2}=\frac{4\pi eck_{B}T(B_{c2}-B)}{1.16 \mu_0 \sigma (2\kappa
_{2}^{2}-1)\Psi   _{2}(\frac{1}{2}+a)},
\end{equation}
  where  $\mu_0$ is  the vacuum  permeability, $\sigma  $   the
  normal  state  conductivity  and   $\Psi  _{2}$  the  first
  derivative of the digamma function. The linear field dependence
  of  the averaged  value of   the energy  gap squared  results in a
  linearity in the above mentioned scaling of the averaged gapless DOS,
  or $G(V)$, with magnetic field.
  By a simple renormalization of the tunneling conductance
  to  their zero-bias  value according to the expression
  $(G(V)-1)/(G  (0)-1)$, the  field-dependent  parameter
  $\bar{\Delta}$ falls out the problem. The in such a way renormalized
  tunneling conductances at a fixed temperature should collapse
  on the same curve for different magnetic fields.
  This behavior can be very useful for the experimental
  determination  of the  pair-breaking parameter  $\alpha$ or
  $ B_{c2}$,  and  of the  order  parameter  $\bar{\Delta},$  or
  $\kappa _{2}$  \cite{degennes1,guyon}. The region of validity of
  expressions  (5),(7)  and  (8) in the $B-T$ diagram   is  not very well  known.
  They are certainly valid in the high-field region $B\rightarrow B_{c2}(T)$.

\vspace{1cm}
\section{Experiment}

  The  high   quality  single-crystalline  Ba$_{1-x}$K$_x$BiO$_3$
  samples  were  grown by  electrochemical
  crystallisation  \cite{crystalgrowth}. The  quality of the
  single  crystals was  verified by  the measurement  of the
resistance in  a four probe  contact configuration. In  zero
magnetic  field  with  decreasing  temperature  the  samples
showed metallic temperature dependence  which saturates to a
residual   resistivity   100   $\mu \Omega$cm   above  the
superconducting transition with a width $\Delta T_c \simeq
1.2$  K.  The  ac-susceptibility  measurement  confirms this
transition width. The critical temperature  $T_c = 23$ K has
been  determined  from  the   midpoint  of  the  zero  field
transition.

  The tunnel junctions were prepared by painting
  a silver spot  on the freshly  cleaned surface of  the crystal.
  The  interface  between  the  silver and Ba$_{1-x}$K$_x$BiO$_3$
  counter electrodes served as a natural barrier forming a planar
  normal metal-insulator-superconductor  (N-I-S) tunnel junction.
  The tunneling measurements were performed in magnetic fields up
  to 30~T perpendicular to the planar tunneling junction enabling
  the  formation of  the vortex  state in  the junction area
\cite{samuely98}.

\vspace{1cm}
\section{Results  and Discussion}

  In Fig.~1 the  experimental tunneling  data are  shown at
  different magnetic fields increasing from zero to 30 Tesla with
  a step  of  2  Tesla  (if  not  otherwise  specified) for three
  temperatures T=1.5, 4.2 and 16 K.
  For a specific temperature the upper critical field can be directly
  determined as the field where  any trace of the  superconducting density
  of states  disappears in the conductance data.
  In our previous analysis of the tunneling data \cite{samuely98}
  we  deduced  the  upper   critical  field  from  a  linear
extrapolation of the magnetic field dependence of the zero-bias conductance.
  The obtained temperature dependence of $B_{c2}$ agreed with the
  standard WHH theory.

  To discuss the validity  of the dirty limit ($l <  \xi_0 $) to our
  system now we  have
  calculated  the BCS  coherence length  $\xi_0 \simeq  49$~{\AA}
  from $B_{c2}(0) = 28$ T
  and take  the mean free path  $l \simeq 33$~{\AA}
    from \cite{affronte}.  The Ginzurg-Landau coherence
  length  is related  to the  BCS one via  $\xi_{GL}=0.855\sqrt{\xi_0
  l}$.  The  estimated  ratio  $l/\xi_0$  is  about $\sim 0.7$.
  One of us (P.W.) has calculated the  density of  states as  a
  function  of  $l/\xi_0$  \cite{Wyder}   and  found  very  small
  differences for the ratio
  $l/\xi_0$   changing  from   0  to   1.  Similarly  Eilenberger
  \cite{Eilenberger} found only small
  differences  for  this  $l/\xi_0$  range  when  calculating the
  temperature  dependence  of  the  Ginzburg  Landau parameters.
  Therefore, we believe  that  the dirty  limit model of
  de  Gennes and  Maki is  applicable in  our case.  The obtained
  agreement  with  experimental  data,  as  discussed further on,
  seems to justify this approach.

  The data  sets as shown in Fig.~1  were  rewritten
  in   the  above  mentioned   form
  $(G(V)-1)/(G(0)-1)$ and are shown in the right side of the Fig.~1.
  This simple renormalization of the conductance reveals,
  that above a  certain magnetic  field all renormalized
  conductance  curves  belonging to the same temperature
  collapse to  the same curve. This  is in a full  agreement with
  Eq.~(5) for the description of the  gapless regime. We  have also
  made the  same rescaling procedure  for the data  sets taken at
  $T$ = 1.5, 3, 4.2, 8, 12, 16  and 20 K. At all temperatures
  an agreement with
  the de Gennes expression can  be found for
  magnetic   fields    $B   >   0.5   B_{c2}(T)$.
  Rewriting Eq.~(5) in the   form $(N(E)-1)/(N(0)-1)$, or
  $(G(V)-1)/(G(0)-1)$, makes that the only unknown  quantity is  the pair-breaking
  parameter $\alpha  $. We have  made a fit  of this renormalized
  gapless DOS to the experimental  data accounting for the thermal
  smearing  as defined  in Eq.~(6). By  this  fitting procedure the same
  $\alpha$  can  be  obtained   for any
  tunneling  trace  at  $  B   >  0.5  B_{c2}(T)$  at  a  particular
  temperature.

  The experimental  values of the  upper critical field  obtained
  directly as mentioned above are displayed in  the Fig.~2
  as a function of temperature together
  with the temperature dependence  of the pair-breaking parameter
  $\alpha$ obtained by  the fits of the voltage dependence of the conductance.
  The
  shown  error bars  account for   the scattering  of the $  \alpha $
  values as  obtained by the  fitting to the  curves at different
  fields. Because the amplitude of the normalized superconducting
  density  of states  decreases with  increasing temperature  the
 error bars increase for increasing temperature.
 Above 16~K the error bars are of the
  same  order   as  $\alpha$  itself.  The direct   relation
  between $\alpha$ and $  B_{c2}$ is defined  by
  Eq.~(3), which depends on  the  diffusion  constant  $D$  of  the  sample.
  The best   agreement between  the experimenatlly determined
  critical field and pair-breaking parameter  is   obtained   for
  $D   =  1$~cm$^2$s$^{-1}$. This  value is in a  perfect agreement with the
  data of  Affronte et al.  \cite{affronte} obtained on a  very
  similar  sample made  in the  same laboratory  and also  agrees
  reasonably  with the  value  of  Roesler et  al. \cite{roesler}
  obtained  for  a  thin  film,  where  they  found  $D  =  0.64$
  cm$^2$s$^{-1}$.

  In  Fig. 3  we show  that the  zero-field tunneling-conductance
  trace at
  1.5 K can  be described by the BCS density  of states using
  the  Dynes formula
  $N(E) \sim {\rm Re}\{ E/(E^2  - \Delta ^2)^{1/2}\}$
   with  the complex  energy  $E = E' +  i\Gamma $
   which takes account for a certain sample inhomogeneity via the
   broadening parameter $\Gamma$.
  This formula gives a good description of the tunnel spectra at 1.5~K and zero magnetic
  field with  $\Delta = 3.9  \pm 0.1$~meV and  $\Gamma = 0.4  \pm
  0.1$~meV, yielding $2\Delta /kT_c = 3.9 \pm 0.1$. This indicates
  that  Ba$_{1-x}$K$_x$BiO$_3$  is a BCS-like  superconductor with
  a medium coupling  strength \cite{huang}.

  In  the case of  weak-coupling
  superconductors,   the
  pair-breaking parameter  at   $T  =  0$~K is
  connected  with  the  energy  gap  $\Delta  (0)$  by the  expression
  $\alpha = \Delta (0)/2$. In our BKBO sample
  at  $T  \rightarrow  0$~K the  superconductivity  is  destroyed at
  $\alpha \sim 2.85  $ meV which is about  $45\%$ higher
    than the expected value
  $\Delta (0)/2  \sim 2$ meV. A similar  discrepancy has been
  found in
tunneling experiments of conventional type-II superconductors
in the dirty limit with a finite value of  $l/\xi_0$  and
with strong-coupling effects
  \cite{guyon}.
  In  our   case  with  $l/\xi_0$  about   $\sim  0.7$  also  the
  strong-coupling  plays  an  important  role  as  shown  in ref.
  \cite{samuely93}.

   In Fig.~3, the  normalized tunneling  conductances $(dI/dV)/(dI/dV)_N$
   at  $T =  1.5$ K are  also
  displayed for magnetic fields $B > 10$~T
where the scaling in the voltage dependence
  of the conductance curves holds. These curves  are
  fitted  by the  de Gennes  formula using the previously determined
  pair-breaking parameter  $\alpha = 2.85$~meV.
Then the only fitting parameter is the superconducting order
parameter $\bar{\Delta}$. Thus, we obtained
  the magnetic-field dependence of the  superconducting  order  parameter
  $\bar{\Delta}$.   The same procedure was repeated
  for the magnetic-field dependent data sets
  at different temperatures.
The results of $\bar{\Delta}(B)$ are  shown in Fig. 4
  for $T = 1.5$, 8, and 16~K. In the inset we display the
  magnetic field dependence of $\bar{\Delta  }^2$.
 $\bar{\Delta}^2(B)$ changes  linearly
  with the applied field not only in the high field region
  near $B_{c2}$ as  predicted by Maki (Eq.~(8))
but at least  from  $B > 0.5 B_{c2}(T)$.
 This finding  supports  the
  evaluation   of  the   upper  critical   field  from  tunneling
  spectroscopy  as  made  in  our  previous  work  where a linear
  extrapolation of the zero-bias tunneling conductance $G(0)$ was
  used   to   obtain   $B_{c2}(T)$ \cite{samuely98}.

  The  normalized
  zero-bias  tunneling conductance $G(V=0,B)_T$ at a certain  temperature
 can be approximated by
\begin{equation}
G(V=0,B)_T=1-P(B)_T(B_{c2}-B),
\end{equation}
  where $P(B)_T$ is the slope of the normalized zero-bias
conductance $G(V=0,B)_T$ versus magnetic field at a fixed
temperature with dimension [T$^{-1}$].
  With this expression   for   the   zero-bias   tunneling  conductance
  substituted  into Eq.~(7)  together with $\bar{\Delta}$
  from  Eq.~(8)  we  can  derive  the  second  GL parameter
  $\kappa _2$ as a function of the temperature \cite{guyon}
\begin{equation}
  \kappa   _{2}(T)    -   0.5   = -    \frac{e}{4.64 \pi \mu_0 k_{B}
T \sigma P(B)_T}\hspace{1mm} \frac{\Psi _3 (\frac{1}{2} + a)}
{\Psi _2 (\frac{1}{2} + a)}.
\end{equation}
  Here,  besides the  slope  $P(B)_T$  of the  zero-bias tunneling
  conductance  and  the
  pair-breaking  parameter $\alpha  $ the  electrical
  conductance   $\sigma$ is an experimental parameter.
For $\sigma$  we    can   take
  $\sigma = 0.125$ x $10^{7}\Omega^{-1}$m$^{-1}$ from   ref.~\cite{affronte}.
  The  resulting   temperature  dependence  of
  $\kappa _2$  is shown in Fig.  5 by open symbols  together with
the   theoretical   prediction  of   Caroli,  Cyrot   and  de  Gennes
  \cite{caroli1}  for a  dirty type-II  s-wave superconductors in
the mixed state.
  Despite the fact  that the error  bars are very
  big  a  discrepancy  is  obvious.  As  shown  by  Guyon  at al.
  \cite{guyon}  the slope  $P(B)_T$ at  different temperatures is
  expected  to be  constant. In   our case  it changes  at higher
  temperatures  (see  inset  in  Fig.  5).  If  the slope is kept
  artificially constant as it was at low temperatures (dotted line
  in the inset) we derive  a much better agreement with the theoretical
  prediction of  Caroli et al. (full  symbols in Fig. 5).
One  possible explanation of this discrepancy  between the theoretical predictions
  of  Caroli et  al. and  our $\kappa  _2(T)$ calculation from the  real
  temperature dependent $P(H)_T$ values
could result from the local sample inhomogeneity
\cite{samuely98,szabo94} because  at higher
  temperatures possibly another  phase can  play a  role
in the tunneling characteristics.
In this case the correct $\kappa  _2  (T)$ dependence of the major bulk
  superconducting phase  is determined from the constant $P(H)_T$
following the predictions of Caroli et al..
 Then,  $\kappa_2(T)$ is equal to
 $\kappa _1(T)$ and it can be used for the determination of basic
  superconducting quantities, connected
with the GL parameters, like the thermodynamic critical  magnetic
field or the free energy.

Anyhow, the $\kappa _2 (T)$ parameter
  calculated from our magnetotunneling data reveals a saturation
 at  low temperatures.
  Its  value  at  higher temperatures is in
  a qualitative  agreement with  the earlier  published values of
  the  GL parameter  $\kappa $  in the  BKBO system \cite{kappa}.
The saturating  character of  $\kappa _2  (T)$  in the low
  temperature  region   proves a dirty limit  and a non  $d$-wave
  character of the superconductivity in the BKBO system.
  Pure  superconductors \cite{Eilenberger} similarly  as  $d$-wave  superconductors
  should   reveal   a  characteristic   $\kappa  _2   (T)  \sim \ln(1/T)$
  dependence \cite{makipriv}.

\vspace{1cm}
\section{Conclusions}

  The pair-breaking theory  has been applied  for
  analyzing   the   magnetotunneling    results   in   the   BKBO
  superconductor. It has been  shown, that the generalization of
  the Ginzburg-Landau  theory provided by Maki  and de Gennes for
  the  dirty type-II  superconductors in  a field  region close to
  $B_{c2}$ is valid in a much wider interval of magnetic fields.
The theory  for the gapless
  superconductors describes the experimental tunneling data
for magnetic fields  $B >  0.5 B_{c2}(T)$ very well.
In  this region of
  magnetic  fields  the  tunneling  curves can be fitted
  by  the de Gennes  formula for the thermally  smeared
  averaged density of states.
As  a  consequence  the  tunneling  curves  reveal an
universal scaling in the magnetic field dependence for a fixed
temperature.
This allows  us to calculate the temperature
  dependence of the pair-breaking
  parameter  $\alpha$  (or $B_{c2}$)  and  the Maki parameter
  $\kappa _2$ from the tunneling curves.
  The   averaged   superconducting    order   parameter   squared
  $\bar{\Delta}^2$
  reveals linearity with the magnetic  field in the same range of
  magnetic fields.
  This linearity can  be applied not only in interpretation of
  the tunneling measurements  but also in the case of
  many superconducting quantities governed by $\bar{\Delta}^2(H)$
in the high-field limit,   like   magnetization, magnetic
  susceptibility, thermal conductivity,  etc..
  The obtained temperature
  dependence  of the  upper critical  field is  in satisfactory
  agreement with the  theoretical
  predictions  of  Maki   and  Werthamer, Helfand and
  Hohenberg for classical  type-II superconductors.

  The wide  magnetic field range  where the pair-breaking  theory
  has  been proved  to be   valid (far  below the  real $B_{c2}$)
makes this tunneling approach very promising
for the high-$T_c$ cuprates
  where the upper critical  field is experimentally not available
for reliable estimates  of  the temperature  dependence of $B_{c2}$.

\acknowledgments

We greatly acknowledge  stimulating discussions with  K. Maki.
This work has been supported
by the EC grant No.CIPA-CT93-0183  and the Slovak
VEGA contract No. 2/5144/98.


\vskip 2 in

\newpage
{\bf FIGURE CAPTIONS}

Fig. 1:
  {Normalized ($G$, left side) and renormalized
  ($(G - 1)/(G(0)-1)$, right side) tunneling
  conductances  of the  Ba$_{1-x}$K$_x$BiO$_3$-Ag tunnel junction
  in magnetic fields from zero up to 30 T in steps of 2~T (if not
  mentioned  else) at  the indicated  temperatures.
The renormalized curves   have been fitted  to the de Gennes
  formula (Eq.~(5)) of the  DOS (open  circles) yielding
the indicated pair-breaking parameter $\alpha$.\\}

Fig. 2:
  {Temperature dependence of the upper critical magnetic field
$B_{c2}$ of  Ba$_{1-x}$K$_x$BiO$_3$  determined directly  from
an extrapolation of the zero-bias tunneling conductance
(closed  circles, left  scale)  and
of  the pair-breaking  parameter $\alpha$ determined
from a fit to the de Gennes formula of the DOS
for the full voltage-dependence of the conductance
(open squares, right  scale).
The  full  line  shows  the prediction  of the standard WHH theory.\\}

Fig. 3:
  {Experimentally  measured normalized  tunneling conductances  at
  $T = 1.5$ K (closed squares) shifted along the Y-axis as
indicated by the high-voltage values at the right side.
The lines show  the BCS dependence at $B = 0$  T (dotted line)
and the fitting to the de Gennes pair-breaking (PB) formula of the DOS
at $B > 10$ T (full lines)\\}.

Fig. 4:
  {Magnetic field dependencies of the averaged value of the energy gap
  $\bar{\Delta}$ at  different temperatures. The  inset shows the
  squared values $\bar{\Delta}^2$ as a function of magnetic field.\\}

Fig. 5:
  {Temperature dependence of the second Ginzburg-Landau parameter
  $\kappa  _2$  calculated  from  magnetotunneling  results (left
  scale) for the temperature dependent $P(B)_T$  (open circles)
and for a constant $P(B)_T= 0.0395$ T$^{-1}$ (closed squares).
The full line   shows  the  theoretical   prediction   of   Caroli  et  al.
  \cite{caroli1}. The  inset displays the temperature  dependence of
  the zero-bias slope $P(B)_T$. }

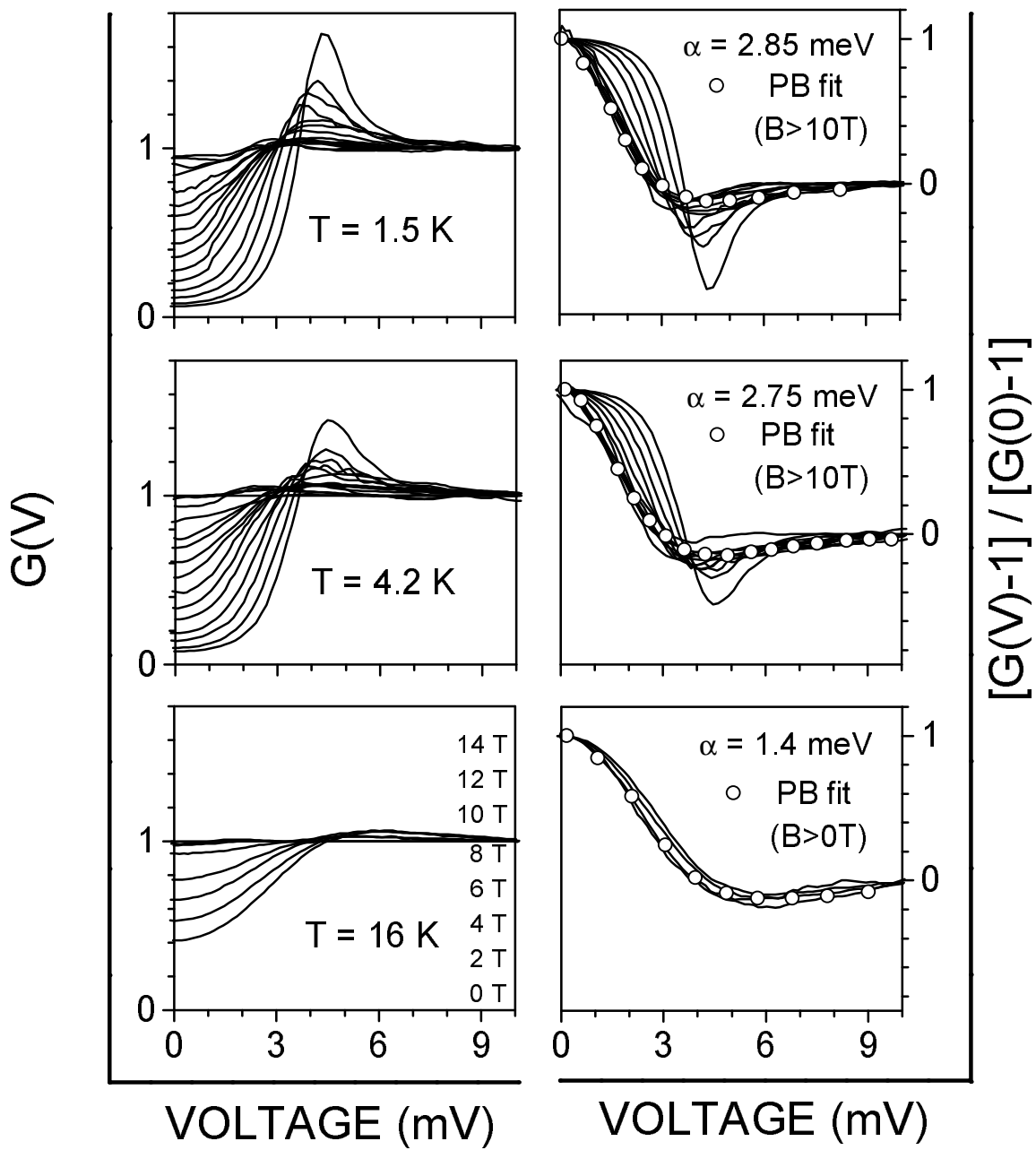
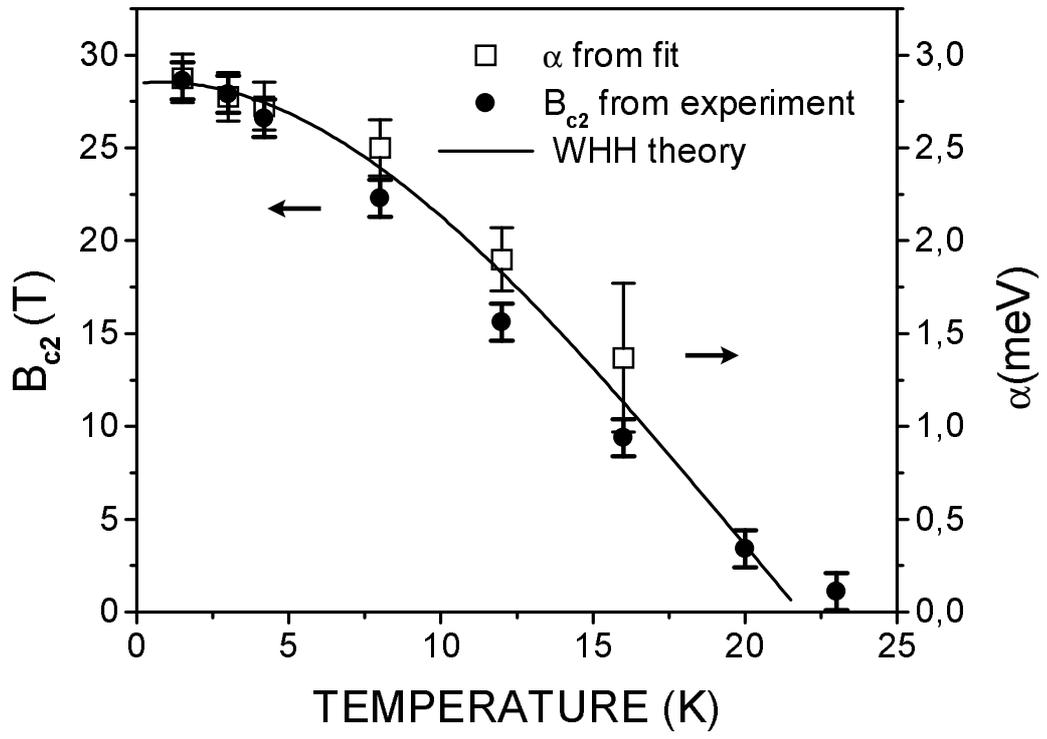
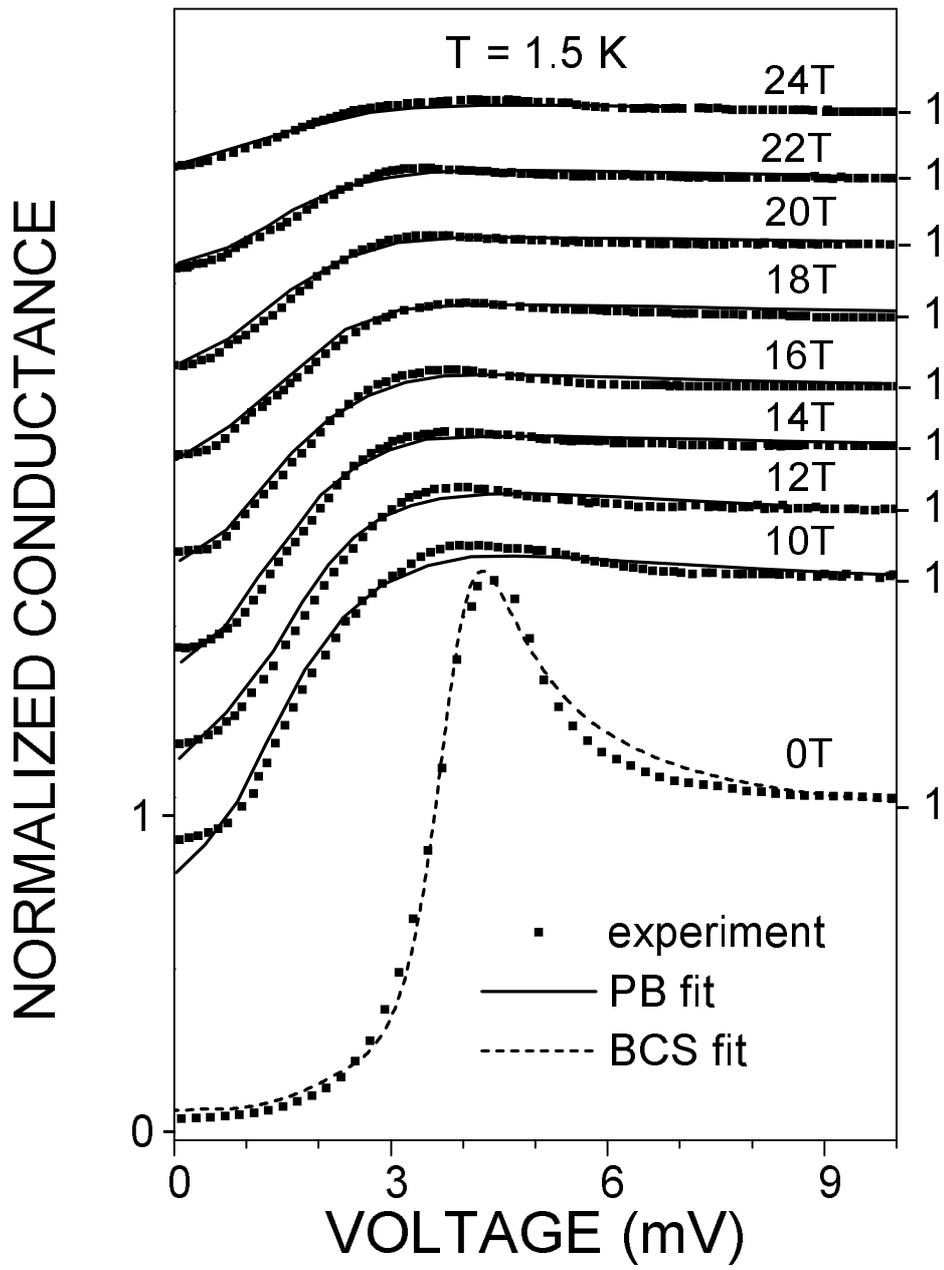
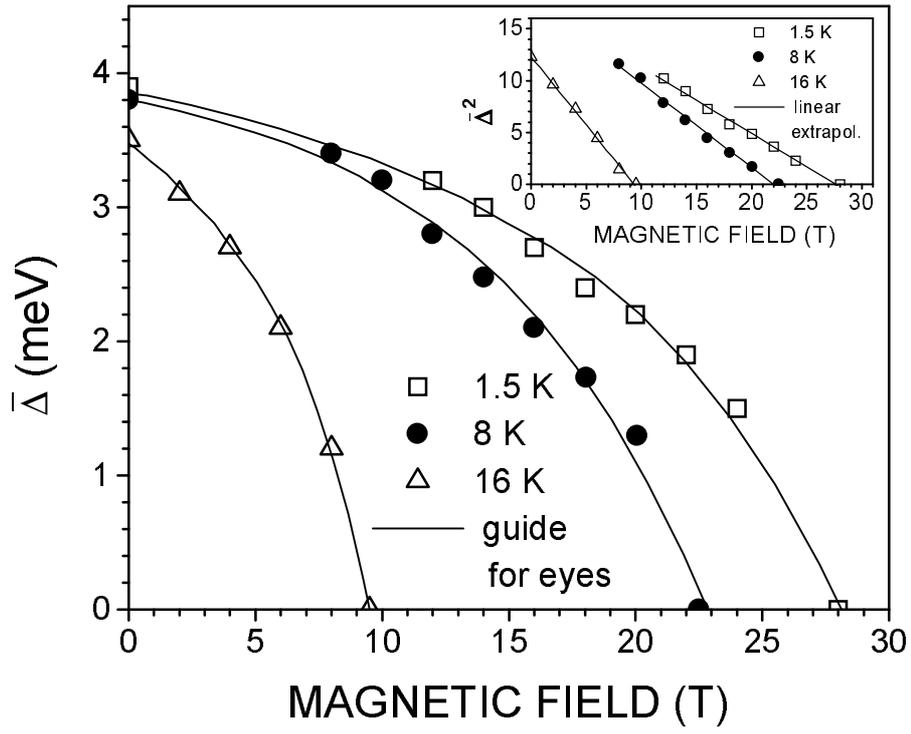
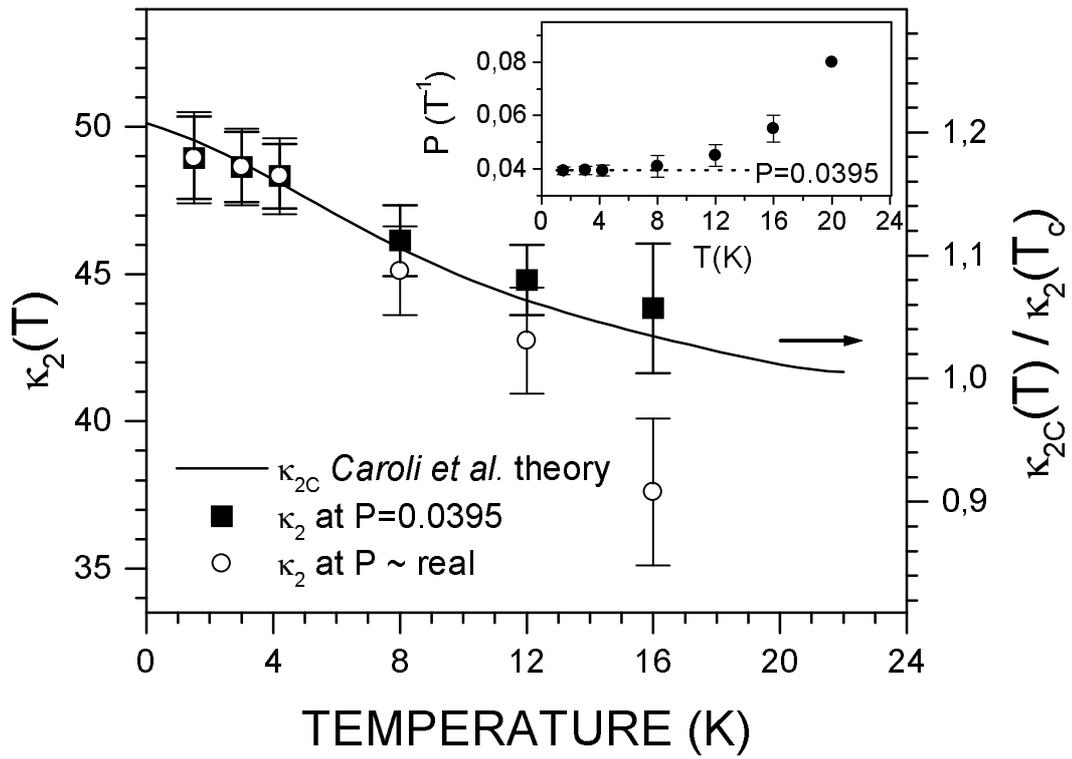
\end{document}